\pdfoutput=1
\documentclass[]{osa-article}

\journal{osajournal}

\articletype{Research Article}

\usepackage{empheq}
\usepackage{bm}
\usepackage[normalem]{ulem}

\let\originalleft\left
\let\originalright\right
\renewcommand{\left}{\mathopen{}\mathclose\bgroup\originalleft}
\renewcommand{\right}{\aftergroup\egroup\originalright}
\DeclareMathAlphabet{\mathcal}{OMS}{cmsy}{m}{n}

\providecommand{\mb}[1]{\mathbf{#1}}
\providecommand{\msf}[1]{\mathsf{#1}}

\providecommand{\mc}[1]{\mathcal{#1}}
\providecommand{\ro}{\mathbf{\mathbf{r}}_o}
\providecommand{\so}{\mathbf{\hat{s}}_o}

\providecommand{\rbm}[1]{r_b^{\text{m}}}
\providecommand{\rd}{\mathbf{r}_d}

\providecommand{\mh}[1]{\mathbf{\hat{#1}}}
\providecommand{\mbb}[1]{\mathbb{#1}}
\providecommand{\bs}[1]{\boldsymbol{#1}}
\providecommand{\bv}{\bs{\nu}}
\providecommand{\taup}{\bs{\tau}}

\providecommand{\lmsum}{\sum_{\ell=0}^\infty\sum_{m=-\ell}^{\ell}}

\providecommand{\ints}[1]{\int_{\mbb{S}^{#1}}}

\begin{document}

\title{Spatio-angular fluorescence microscopy\\ I. Basic theory}

\author{Talon Chandler,\authormark{1,*} Hari Shroff,\authormark{2,3} Rudolf Oldenbourg,\authormark{3} and Patrick La Rivi\`ere\authormark{1,3}}

\address{\authormark{1}University of Chicago, Department of Radiology, Chicago, Illinois 60637, USA\\
  \authormark{2}Section on High Resolution Optical Imaging, National Institute
  of Biomedical Imaging and Bioengineering, National Institutes of Health,
  Bethesda, Maryland 20892, USA\\
  \authormark{3}Marine Biological Laboratory, Bell Center, Woods Hole, Massachusetts 02543, USA
}

\email{\authormark{*}talonchandler@talonchandler.com} 



\begin{abstract*}
  We introduce the basic elements of a spatio-angular theory of fluorescence
  microscopy, providing a unified framework for analyzing systems that image
  single fluorescent dipoles and ensembles of overlapping dipoles that label
  biological molecules. We model an aplanatic microscope imaging an ensemble of
  in-focus fluorescent dipoles as a linear Hilbert-space operator, and we show
  that the operator takes a particularly convenient form when expressed in a
  basis of complex exponentials and spherical harmonics---a form we call the
  dipole spatio-angular transfer function. We discuss the implications of our
  analysis for all quantitative fluorescence microscopy studies and lay out a
  path towards a complete theory. 
\end{abstract*}

\section{Introduction}
Fluorescence microscopes are widely used in the biological sciences for imaging
fluorescent molecules that label specific proteins and biologically important
molecules. While most fluorescence microscopy experiments are designed to
measure only the spatial distribution of fluorophores, a growing number of
experiments seek to measure both the spatial and angular distributions of
fluorophores by the use of polarizers \cite{vrabioiu2006, mattheyses2010,
  mehta2016, mcquilken2017, zhanghao2017} or point spread function engineering
\cite{agrawal2012, zhang2018}.

Meanwhile, single-molecule localization microscopy (SMLM) experiments use
spatially sparse fluorescent samples to localize single molecules with precision
that surpasses the diffraction limit. Noise limits the precision of this
localization \cite{foreman2011, chao2016}, and several studies have shown that
model mismatch (e.g. ignoring the effects of vector optics, dipole orientation
\cite{backlund2014}, and dipole rotation \cite{lew2013}) can introduce
localization bias as well. Therefore, the most precise and accurate SMLM
experiments must use an appropriate model and jointly estimate both the position
and orientation of each fluorophore. Several studies have successfully used
vector optics and dipole models to estimate the position and orientation of
single molecules \cite{bohmer2003, lieb2004, toprak2006, aguet2009,
  mortensen2010}, and there is growing interest in designing optical systems for
measuring the position, orientation, and rotational dynamics of single molecules
\cite{agrawal2012, backer2014, stallinga2015, zhang2018, zhang2018-2}.

While many studies have focused on improving imaging models for spatially sparse
fluorescent samples, we consider the more general case and aim to improve
imaging models for arbitrary samples including those containing ensembles of
fluorophores within a resolvable volume. In particular, we examine the effects
of two widely used approximations in fluorescence microscopy---the
\textit{monopole approximation} and the \textit{scalar approximation}.

We use the term \textit{monopole} to refer to a model of a fluorophore that
treats it as an isotropic absorber/emitter. Although the term \textit{monopole
  approximation} is not in widespread use, we think it accurately describes the
way many models of fluorescence microscopy treat fluorophores, and we use the
term to distinguish the monopole model from more realistic dipole and
higher-order models. Despite their use in models, electromagnetic monopole
absorber/emitters do not exist in nature. All physical fluorophores absorb and
emit radiation with dipole or higher-order moments, and these moments are always
oriented in space. For a classical mental model of fluorophores we imagine each
dipole as a small oriented antenna (with incoherent absorption and emission
moments) where electrons are constrained to move along a single direction. 

All fluorescence microscopy models that use an \textit{optical point spread
  function} or an \textit{optical transfer function} to describe the mapping
between the fluorophores and the irradiance on the detector implicitly make the
monopole approximation. The optical point spread function is the irradiance
response of an optical system to an isotropic point source, so it cannot model
the response due to an anisotropic dipole radiator. In this work we define
\textit{monopole and dipole transfer functions} that describe the mapping
between fluorophores and the measured irradiance. Although optical systems are
an essential part of microscopes, fluorescence microscopists are interested in
measuring the properties of fluorophores (not optics), so the monopole and
dipole transfer functions are more directly useful than the optical transfer
function for the problems that fluorescence microscopists are interested in
solving.

While the monopole approximation applies to the fluorescent object, the
\textit{scalar approximation} applies to the fields that propagate through the
microscope. Modeling the electric fields in a region requires a
three-dimensional vector field, but if the electric fields are random or
completely parallel, a scalar field is sufficient, and we can replace the
vector-valued electric field, $\mb{E}$, with a scalar-valued field, $U$.

The scalar approximation is often made together with the monopole approximation.
For example, the Born-Wolf model \cite{born1980} and the Gibson-Lanni model
\cite{gibson89} make both the monopole and scalar approximations when applied to
fluorescence microscopes. However, some models make the monopole approximation
but not the scalar approximation. For example, the Richards-Wolf model
\cite{richards} considers the role of vector-valued fields in the optical
system, but it is an optical model so when it is applied to fluorescence
microscopes the monopole approximation is assumed. 

This work lies at the intersection of three subfields of fluorescence
microscopy: (1) spatial ensemble imaging where each resolvable volume contains
many fluorophores and the goal is to find the concentration of fluorophores as a
function of position in the sample, (2) spatio-angular ensemble imaging where
each resolvable volume contains many fluorophores and the goal is to find the
concentration and average orientation of fluorophores as a function of position
in the sample, and (3) SMLM imaging where fluorophores are sparse in the sample
and the goal is to find the position and orientation of each fluorophore. We
briefly review how these three subfields use the monopole and scalar
approximations.

The large majority of fluorescence microscopes are used to image ensembles of
fluorophores, and most existing modeling techniques make use of the monopole
approximation, the scalar approximation, or both. As discussed above, the
Gibson-Lanni model, the Born-Wolf model, and the Richards-Wolf model are
approximate when applied to fluorescence microscopy data because they only model
monopole emitters. Deconvolution algorithms that use these models may make
biased estimates of fluorophore concentrations since they ignore the dipole
excitation and emission of fluorophores. 

A small but growing group of microscopists is interested in measuring the
orientation and position of ensembles of fluorophores \cite{vrabioiu2006,
  mattheyses2010, mehta2016, mcquilken2017, zhanghao2017}. These techniques
typically use polarizers to make multiple measurements of the same object with
different polarizer orientations, then they use a model of the dipole excitation
and emission processes \cite{fourkas2001} to recover the orientation of
fluorophores using pixel-wise arithmetic. Although these studies do not adopt
the scalar or monopole approximations for the angular part of the problem, they
adopt both approximations when they consider the spatial part of the problem.
Existing works either ignore the spatial reconstruction problem
\cite{vrabioiu2006, mattheyses2010, mehta2016, mcquilken2017} or assume that the
spatial and angular reconstruction problems can be solved sequentially
\cite{zhanghao2017}.

The most precise experiments in SMLM imaging do not adopt the scalar or monopole
approximations. Although many works have applied dipole models with vector
optics, fewer have considered the effects of rotational or spatial diffusion,
and to our knowledge no works have considered both rotational and spatial
diffusion together. We will see that the dipole transfer functions are useful
tools for incorporating angular and spatial diffusion into SMLM simulations and
reconstructions.

In the present work we begin to place these three subfields on a common
theoretical footing. First, in section \ref{sec:theory}, we consider arbitrary
fluorescence imaging models and lay out a plan for developing a model for
spatio-angular imaging. In section \ref{sec:monopole} we review the familiar
monopole imaging model, and in section \ref{sec:dipole} we extend the model to
dipoles. Finally, in section \ref{sec:discussion} we discuss the results and
their broader implications.

In this paper we focus on modeling a single-view fluorescence microscope without
polarizers. In future papers of this series we will extend our models to include
polarizers and multi-view microscopes. Additionally, we have restricted this
paper to the forward problem---the mapping between a known object and the data. In
future papers we will consider the inverse problem, and the singular value
decomposition (SVD) will play a central role.
 
\section{Theory}\label{sec:theory}
We begin our analysis with the abstract Hilbert space formalism of Barrett and
Myers \cite[ch.~1.3]{barrett2004}. Our first task is to formulate the imaging
process as a mapping between two Hilbert spaces
$\mc{H}: \mbb{U} \rightarrow \mbb{V}$, where $\mbb{U}$ is a set that contains
all possible objects, $\mbb{V}$ is a set that contains all (possibly
noise-corrupted) datasets, and $\mc{H}$ is a model of the instrument that maps
between these two spaces. We denote (possibly infinite-dimensional)
Hilbert-space vectors in $\mbb{U}$ with $\mb{f}$, Hilbert-space vectors in
$\mbb{V}$ with $\mb{g}$, and the mapping between the spaces with
\begin{align}
  \mb{g} = \mc{H}\mb{f}.
\end{align}
Throughout this work we will use the letters $g$, $h$, and $f$ with varying
fonts, capitalizations, and arguments to represent the data, the instrument, and
the object, respectively.

Once we have identified the spaces $\mbb{U}$ and $\mbb{V}$, we can start
expressing the mapping between the spaces in a specific object-space and
data-space basis. In most cases the easiest mapping to find uses a
delta-function basis---we expand object and data space into delta functions,
then express the mapping as an integral transform. After finding this mapping we
can start to investigate the same mapping in different bases.

The above discussion is quite abstract, but it is a powerful point of view that
will enable us to unify the analysis of spatio-angular fluorescence imaging.
In section \ref{sec:monopole} we will demonstrate the formalism by examining a
familiar monopole imaging model, and we will demonstrate the mapping between
object and data space in two different bases. In section \ref{sec:dipole} we
will extend the monopole imaging model to dipoles and examine the mapping in
four different bases.

\section{Monopole imaging}\label{sec:monopole}
We start by considering a microscope that images a field of in-focus monopoles
by recording the irradiance on a two-dimensional detector. This section treads
familiar ground, but it serves to establish the concepts and notation that will
be necessary when we extend to the dipole case.

We can represent the object as a function that assigns a real number to each
point on a plane, so we identify object space as
$\mbb{U} = \mbb{L}_2(\mbb{R}^2)$---the set of square-integrable functions on the
two-dimensional plane. Similarly, we have a two-dimensional detector that
measures a real number at each point on a plane, so data space is the same set
$\mbb{V} = \mbb{L}_2(\mbb{R}^2)$.

Next, we name the representations of our object and data in a specific basis. In
a delta function basis the object can be represented by a function $f(\ro)$
called the \textit{monopole density}---the number of monopoles per unit area at
the two-dimensional position $\ro$. Similarly, in a delta function basis the
data can be represented by a function $g'(\rd')$ called the
\textit{irradiance}---the power received by a surface per unit area at position
$\rd'$. Note that we have adopted a slightly unusual convention of using primes
to denote unscaled coordinates. Later in this section we will introduce unprimed
scaled coordinates that we will use throughout the rest of the
paper.

A reasonable starting point is to assume that the relationship between the
object and the data is \textit{linear}---this is true in many fluorescence
microscopes because fluorophores emit incoherently, so a scaled sum of
fluorophores will result in a scaled sum of the irradiance patterns created by
the individual fluorophores. Note that our assumption of linearity excludes
cases where fluorophores interact (e.g. homoFRET) or saturate (e.g. non-linear
fluorescence microscopy).

If the mapping is linear, we can write the irradiance as a weighted integral
over a field of monopoles
\begin{align}
g'(\rd') = \int_{\mbb{R}^2}d\ro\, h'(\rd',\ro)f(\ro), \label{eq:fwdmono}
\end{align}
where $h'(\rd', \ro{})$ is the irradiance at position $\rd'$ created by a
point source at $\ro$.

Next, we assume that the optical system is \textit{aplanatic}---Abbe's sine
condition is satisfied and on-axis points are imaged without aberration. Abbe's
sine condition guarantees that off-axis points are imaged without spherical
aberration or coma \cite[ch.~1]{mansuripur2009}, so the imaging system can
be modeled within the field of view of the optical system as a magnifier with
shift-invariant blur
\begin{align}
  g'(\rd') = \int_{\mbb{R}^2}d\ro\, h'(\rd' - m\ro)f(\ro), \label{eq:nonconv}
\end{align}
where $m$ is a magnification factor. 

We can simplify our analysis by changing coordinates and writing Eq.
\eqref{eq:nonconv} as a convolution \cite[ch.~7.2.7]{barrett2004}. We
define a demagnified detector coordinate $\rd = \rd'/m$ and a normalization
factor that corresponds to the total power incident on the detector plane due to
a point source $P_{\text{mono}} = \int_{\mbb{R}^2}d\mb{r}\,h'(m\mb{r})$ where
$\mb{r} = \rd - \ro$. We use these scaling factors to define the
\textit{monopole point spread function} as
\begin{align}
  h(\rd - \ro) = \frac{h'(m[\rd - \ro])}{P_{\text{mono}}},
\end{align}
and the \textit{scaled irradiance} as
\begin{align}
  g(\rd) = \frac{g'(m\rd)}{P_{\text{mono}}}.
\end{align}
With these definitions we can express the mapping between the object and the
data as a familiar convolution
\begin{align}
  g(\rd) = \int_{\mbb{R}^2}d\ro\, h(\rd - \ro)f(\ro).  \label{eq:lsi}
\end{align}

We have chosen to normalize the monopole point spread function so that
\begin{align}
  \int_{\mbb{R}^2}d\mb{r}\, h(\mb{r}) = 1. \label{eq:norm}
\end{align}
The monopole point spread function corresponds to a measurable irradiance, so it
is always real and positive.

\begin{figure}
  \centering
  \includegraphics[scale=1.0]{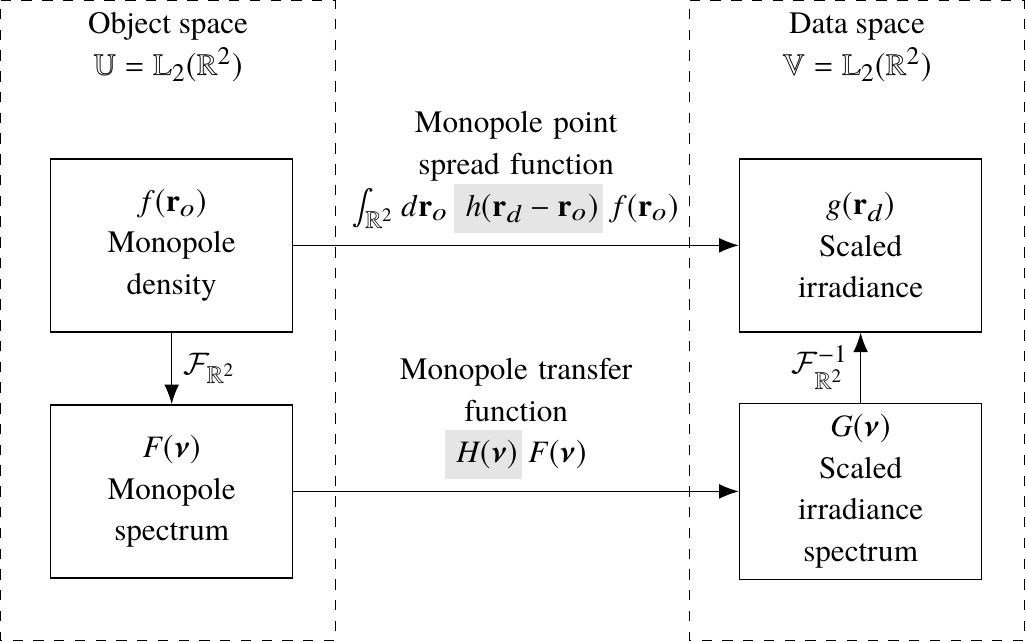}
  \caption{The mapping between the object and data space of a monopole
    fluorescence microscope can be computed in two different bases---a delta
    function basis and a complex exponential basis. The change of basis can be
    computed with a two-dimensional Fourier transform denoted
    $\mathcal{F}_{\mbb{R}^2}$.}
     \label{fig:monopole-block}      
\end{figure}

The mapping between the object and the data in a linear shift-invariant imaging
system takes a particularly simple form in a complex exponential (i.e. Fourier)
basis. If we apply the Fourier convolution theorem to Eq. \eqref{eq:lsi} we find
that
\begin{align}
  G(\bv) = H(\bv)F(\bv),\label{eq:freq}
\end{align}
where we define the \textit{scaled irradiance spectrum} as
\begin{align}
  G(\bv) = \int_{\mbb{R}^2}d\mb{r}\, g(\mb{r})\, \text{exp}(-2\pi i\mb{r}\cdot\bv),
\end{align}
the \textit{monopole transfer function} as
\begin{align}
  H(\bv) = \int_{\mbb{R}^2}d\mb{r}\, h(\mb{r})\, \text{exp}(-2\pi i\mb{r}\cdot\bv),\label{eq:otf}
\end{align}
and the \textit{monopole spectrum} as
\begin{align}
    F(\bv) = \int_{\mbb{R}^2}d\mb{r}\, f(\mb{r})\, \text{exp}(-2\pi i\mb{r}\cdot\bv).
\end{align}
The monopole point spread function is normalized and real, so we know that the
monopole transfer function is normalized, $H(0) = 1$, and conjugate symmetric,
$H(-\bv) = H^*(\bv)$, where $z^*$ denotes the complex conjugate of $z$.

Notice that Eqs. \eqref{eq:lsi} and \eqref{eq:freq} are expressions of the same
mapping between object and data space in different bases. Figure
\ref{fig:monopole-block} summarizes the relationship between object and data
space in both bases.

We have been careful to use the term \textit{monopole transfer function} instead
of the commonly-used term \textit{optical transfer function}. We reserve the
term \textit{optical transfer function} for optical systems---the optical
transfer function maps between an input irradiance spectrum and an output
irradiance spectrum in an optical system. We can use optical transfer functions
to model the propagation of light through a microscope, but ultimately we are
always interested in the object, not the light emitted by the object. We will
find the distinction between the optical transfer function and the object
transfer function to be especially valuable when we consider dipoles in section \ref{sec:dipole}.

\subsection{Monopole coherent transfer functions}
Although the Fourier transform can be used to calculate the monopole transfer
function directly from the monopole point spread function, there is a well-known
alternative that exploits coherent transfer functions. The key idea is that the
monopole point spread function can always be written as the absolute square of a
scalar-valued \textit{monopole coherent spread function}, $c(\rd - \ro)$,
defined by
\begin{align}
  |c(\rd - \ro)|^2 = h(\rd - \ro). \label{eq:absquarescalar}
\end{align}
Physically, the monopole coherent spread function corresponds to the
scalar-valued field on the detector with appropriate scaling.

We can plug Eq. \eqref{eq:absquarescalar} into Eq. \eqref{eq:otf} and use the
autocorrelation theorem to rewrite the monopole transfer function as
\begin{align}
  H(\bv) = \int_{\mbb{R}^2}d\taup\, C(\taup)C^*(\taup - \bv), 
\end{align}
where we have introduced the
\textit{monopole coherent transfer function} as the two-dimensional Fourier
transform of the monopole coherent spread function:
\begin{align}
  C(\taup) = \int_{\mbb{R}^2}d\mb{r}\, c(\mb{r})\,\text{exp}[-2\pi i\mb{r}\cdot\taup].
\end{align}
Physically, the monopole coherent transfer function corresponds to the
scalar-valued field in a Fourier plane of the detector with appropriate scaling.

\begin{figure}
  \centering
  \includegraphics[scale=1.0]{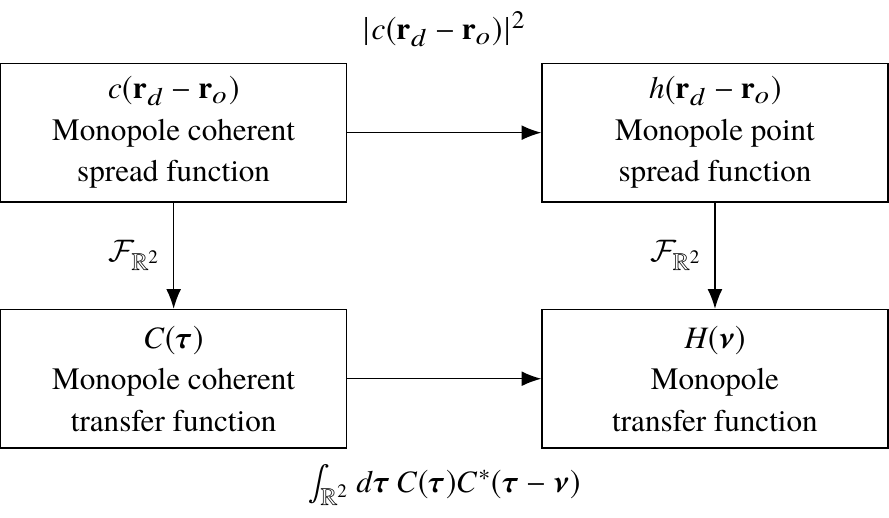}
  \caption{The monopole transfer functions are related by a
    two-dimensional Fourier transform (right column). The coherent monopole
    transfer functions (left column) can be used to simplify the calculation of
    the remaining transfer functions.}
   \label{fig:monopole-transfer-functions}
 \end{figure}

The coherent transfer function provides a valuable shortcut for analyzing
microscopes since it is often straightforward to calculate the field in a
Fourier plane of the detector. A typical approach for calculating the transfer
functions is to (1) calculate the field in a Fourier plane of the detector, (2)
scale the field to find the monopole coherent transfer function, then (3) use
the relationships in Fig. \ref{fig:monopole-transfer-functions} to calculate
the other transfer functions.

\section{Dipole imaging}\label{sec:dipole}
Now we consider a microscope imaging a field of in-focus dipoles by recording
the irradiance on a two-dimensional detector. A function that assigns a real
number to each point on a plane is not sufficient to specify a field of dipoles
because the dipoles can have different orientations. To represent the object we
need to extend object space to
$\mbb{U} = \mbb{L}_2(\mbb{R}^2\times\mbb{S}^2)$---the set of square-integrable
functions on the product space of a plane and a two-dimensional sphere (the
usual sphere embedded in $\mbb{R}^3$). To visualize functions in object space we
imagine a sphere at every point on a plane with a scalar value assigned to every
point on each sphere.

In a delta function basis the object can be represented by a function
$f(\ro, \so)$ called the \textit{dipole density}---the number of dipoles per
unit area and per unit solid angle at position $\ro{}$ and oriented along
$\so{}$. Similar to the monopole case, we model the mapping between the object
and the irradiance in a delta function basis as an integral transform
\begin{align}
  g'(\rd') = \int_{\mbb{S}^2}d\so\int_{\mbb{R}^2}d\ro\, h'(\rd', \ro, \so)f(\ro, \so),
\end{align}
where $h'(\rd', \ro, \so)$ is the irradiance at position $\rd'$ created by a
point source at $\ro$ with orientation $\so$. Notice that we have considered all
possible orientations $\so$ and integrated over the sphere $\mbb{S}^2$. The
dipole density is always symmetric under angular inversion,
$f(\ro, \so) = f(\ro,-\so)$, so we could have chosen to integrate over a
hemisphere and adjusted the definition of the dipole density by a factor of two.
For convenience we will continue to integrate over the complete sphere. We note
that all functions in this work with $\so$ as an independent variable are
symmetric under angular inversion, $\so \rightarrow -\so$.

If the optical system is aplanatic, we can write the
integral transform as
\begin{align}
  g'(\rd') = \int_{\mbb{S}^2}d\so\int_{\mbb{R}^2}d\ro\, h'(\rd' - m\ro, \so)f(\ro, \so). 
\end{align}

We define the same demagnified detector coordinate $\rd = \rd'/m$ and a new
normalization factor that corresponds to the total power incident on the
detector due to a spatial point source with an angularly uniform distribution of
dipoles
$P_\text{dip} = \int_{\mbb{S}^2}d\so\int_{\mbb{R}^2}d\mb{r}\, h'(m\mb{r},
\so)$. We use these scaling factors to define the
\textit{dipole point spread function} as
\begin{align}
  h(\rd - \ro, \so) = \frac{h'(m[\rd - \ro], \so)}{P_\text{dip}}, 
\end{align}
and the \textit{scaled irradiance} as 
\begin{align}
  g(\rd) = \frac{g'(m\rd)}{P_\text{dip}}. 
\end{align}
With these definitions we can express the mapping between the object and the
data as
\begin{align}
g(\rd{}) = \int_{\mbb{S}^2}d\so{}\int_{\mbb{R}^2}d\ro{}\, h(\rd{} -\ro{}, \so{})f(\ro, \so). \label{eq:odpsf}
\end{align}
Equation \eqref{eq:odpsf} is a key result because it represents the mapping
between object space and data space in a delta function basis. The integrals in
Eq. \eqref{eq:odpsf} would be extremely expensive to compute for an arbitrary
object, but the integrals simplify to an efficient sum if the object is
spatially and angularly sparse. In other words, Eq. \eqref{eq:odpsf} is ideal for
simulating and analyzing single fluorophores that are rigidly attached to an
oriented structure.

Similar to the monopole case, we have chosen to normalize the dipole point
spread function so that
\begin{align}
  \int_{\mbb{S}^2}d\so\int_{\mbb{R}^2}d\mb{r}\, h(\mb{r}, \so) = 1. 
\end{align}
The dipole point spread function is a measurable quantity, so it is real
and positive.

\subsection{Dipole spatial transfer function}
We can make our first change of basis by applying the Fourier-convolution
theorem to Eq. \eqref{eq:odpsf}, which yields
\begin{align}
G(\bv) = \int_{\mbb{S}^2}d\so\, H(\bv, \so)F(\bv, \so) \label{eq:odotf},
\end{align}
where we define the \textit{dipole spatial transfer function} as
  \begin{align}
  H(\bv, \so) &= \int_{\mbb{R}^2}d\mb{r}\, h(\mb{r}, \so)\, \text{exp}(-2\pi i\mb{r}\cdot\bv),\label{eq:dstf}
  \end{align}
  and the \textit{dipole spatial spectrum} as
  \begin{align}
  F(\bv, \so) &= \int_{\mbb{R}^2}d\mb{r}\, f(\mb{r}, \so)\, \text{exp}(-2\pi i\mb{r}\cdot\bv). 
  \end{align}
  Since the dipole point spread function is normalized and real, we know that
  the dipole spatial transfer function is normalized,
  $\int_{\mbb{S}^2}d\mh{s}\, H(0, \so) = 1$, and conjugate symmetric,
  $H(-\bv, \so) = H^*(\bv, \so)$.
  
  This basis is ideal for simulating and analyzing objects that are angularly
  sparse and spatially dense; e.g. rod-like structures that contain fluorophores
  in a fixed orientation, or rotationally fixed fluorophores that are undergoing
  spatial diffusion.

  \begin{figure}
  \hspace{-5em}
  \includegraphics[scale=1.0]{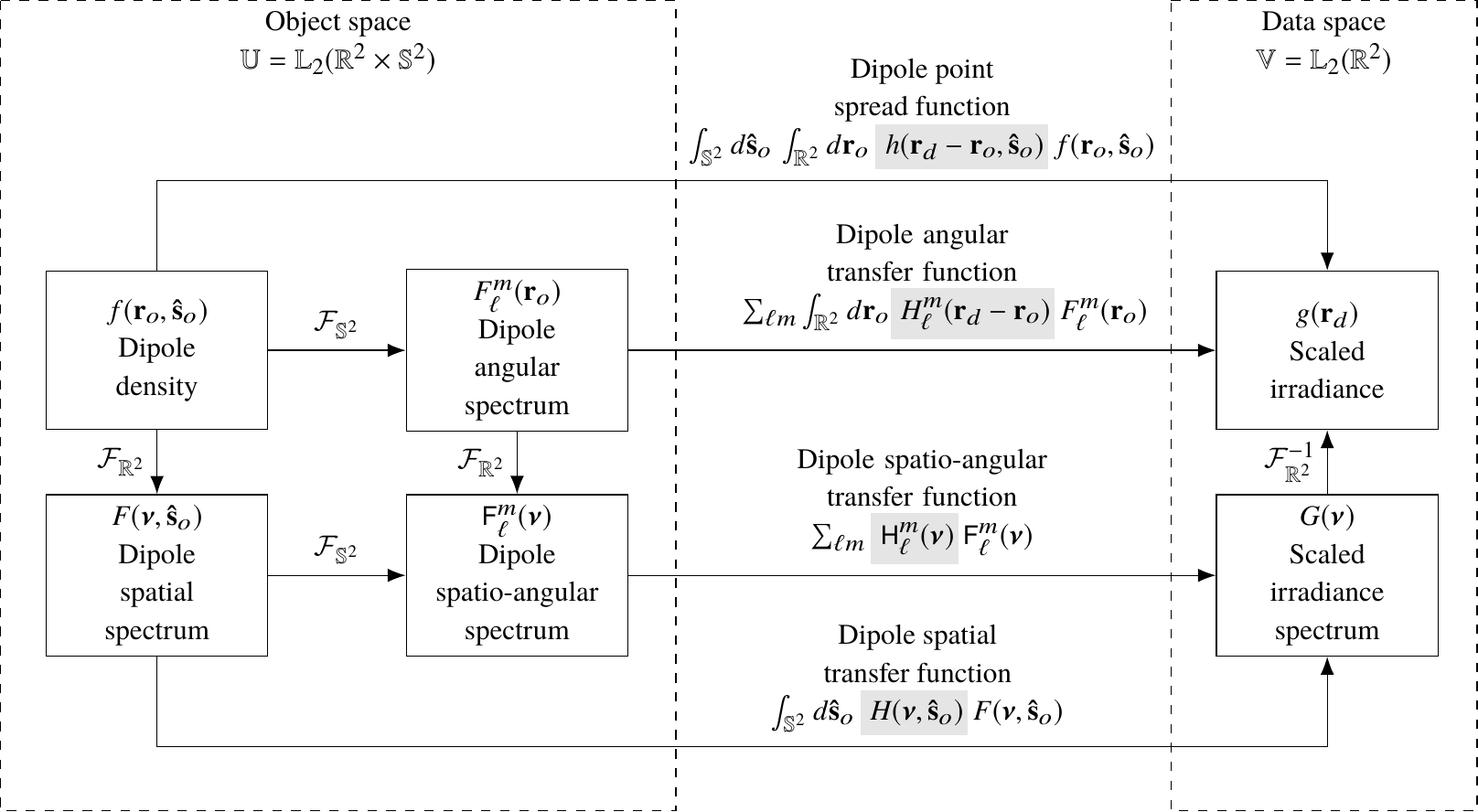}
  \caption{The mapping between the object space and data space of a dipole
    imaging system can be computed in four different bases---a delta function
    basis, a complex-exponential/angular-delta basis, a
    spatial-delta/spherical-harmonic basis, and a
    complex-exponential/spherical-harmonic basis. The changes of basis can be
    computed with the two-dimensional Fourier transform denoted
    $\mathcal{F}_{\mbb{R}^2}$, and the spherical Fourier transform denoted
    $\mathcal{F}_{\mbb{S}^2}$.}
   \label{fig:dipole-block}      
    \end{figure}

\subsection{Dipole angular transfer function}
The spherical harmonics are another set of convenient basis functions that play
the same role as complex exponentials in spatial transfer functions---see
Appendix \ref{sec:sph} for an introduction to the spherical harmonics. We can
change basis from spherical delta functions to spherical harmonics by applying
the generalized Plancherel theorem for spherical functions
\begin{align}
  \ints{2}d\mh{s}\, p(\mh{s})q^*(\mh{s}) = \lmsum P_\ell^m Q_\ell^{m*}, \label{eq:plan}
\end{align}
where $p(\mh{s})$ and $q(\mh{s})$ are arbitrary functions on the sphere,
$P_\ell^m$ and $Q_\ell^m$ are their spherical Fourier transforms defined by
\begin{align}
  P_\ell^m = \int_{\mbb{S}^2}d\mh{s}\, p(\mh{s})Y_\ell^{m*}(\mh{s}),
\end{align}
and $Y_{\ell}^m(\mh{s})$ are the spherical harmonic functions defined in
Appendix \ref{sec:sph}. Equation \eqref{eq:plan} expresses the fact that scalar
products are invariant under a change of basis \cite[Eq.~3.78]{barrett2004}.
The left-hand side of Eq. \eqref{eq:plan} is the scalar product of
$\mbb{L}_2(\mbb{S}^2)$ functions in a delta function basis and the right-hand
side is the scalar product of $\mbb{L}_2(\mbb{S}^2)$ functions in a spherical
harmonic function basis. Applying Eq. \eqref{eq:plan} to Eq. \eqref{eq:odpsf} yields
\begin{align}
  g(\rd) = \lmsum \int_{\mbb{R}^2}d\ro\, H_\ell^m(\rd - \ro)F_\ell^m(\ro), \label{eq:atf-form}
\end{align}
where we have defined the \textit{dipole angular transfer function} as
\begin{align}
  H_\ell^m(\rd - \ro) = \int_{\mbb{S}^2}d\so\, h(\rd - \ro, \so)Y_{\ell}^{m*}(\so),\label{eq:atf-prep} 
\end{align}
and the \textit{dipole angular spectrum} as
\begin{align}
  F_\ell^m(\ro) = \int_{\mbb{S}^2}d\so\, f(\ro, \so)Y_{\ell}^{m*}(\so).
\end{align}
Since the dipole point spread function is normalized and real, we know
that the dipole angular transfer function is normalized,
$\int_{\mbb{R}^2}d\mb{r}\, H_0^0(\mb{r}) = 1$, and conjugate symmetric,
$H_\ell^{-m}(\mb{r}) = (-1)^mH_\ell^{m*}(\mb{r})$.

This basis is well suited for simulating and analyzing objects that are
spatially sparse and angularly dense; e.g. single fluorophores that are
undergoing angular diffusion, or many fluorophores that are within a resolvable
volume with varying orientations.

\subsection{Spatio-angular dipole transfer function}
We can arrive at our final basis in two ways: by applying the generalized
Plancherel theorem for spherical functions to Eq. \eqref{eq:odotf} or by applying
the Fourier convolution theorem to Eq. \eqref{eq:atf-form}. We follow the
first path and find that
\begin{align}
G(\bv) = \lmsum \msf{H}_\ell^m(\bv)\msf{F}_\ell^m(\bv) \label{eq:saft},
\end{align}
where we have defined the \textit{dipole spatio-angular transfer function} as
  \begin{align}
  \msf{H}_\ell^m(\bv) &= \int_{\mbb{S}^2}d\so\, H(\bv, \so)Y_\ell^{m*}(\so),
  \end{align}
  and the \textit{dipole spatio-angular spectrum} as
  \begin{align}
  \msf{F}_\ell^m(\bv) &= \int_{\mbb{S}^2}d\so\, F(\bv, \so)Y_\ell^{m*}(\so).
  \end{align}
  Since the dipole point spread function is normalized and real, we know
  that the dipole spatio-angular transfer function is normalized,
  $\msf{H}_0^0(0) = 1$, and conjugate symmetric,
  $\msf{H}_\ell^{-m}(-\bv) = (-1)^m\msf{H}_\ell^{m*}(\bv)$.
  
  This basis is well suited for simulating and analyzing arbitrary samples
  because it exploits the band limit of the imaging system. We note that most
  single molecule imaging experiments are best described in this basis because
  of the effects of spatial and rotational diffusion.

  Figure \ref{fig:dipole-block} summarizes the relationships between the four
  bases that we can use to compute the image of a field of dipoles. We reiterate
  that all four bases may be useful depending on the sample.
    
\subsection{Dipole coherent transfer functions}
Similar to the monopole case, there is an efficient way to calculate the
transfer functions using coherent transfer functions. The dipole point spread
function can always be written as the absolute square of a vector-valued
function, $\mb{c}(\rd - \ro, \so)$, called the \textit{dipole coherent spread
  function}:
\begin{align}
  |\mb{c}(\rd - \ro, \so)|^2 = h(\rd - \ro, \so). \label{eq:absquare2}
\end{align}
Physically, the dipole coherent spread function corresponds to the vector-valued
electric field on the detector with appropriate scaling. We need a vector-valued
coherent transfer function since the polarization of the field plays a
significant role in dipole imaging, so the dipole point spread function cannot
be written as an absolute square of a scalar-valued function.

\begin{figure}
  \hspace{-2em}
  \includegraphics[scale=1.0]{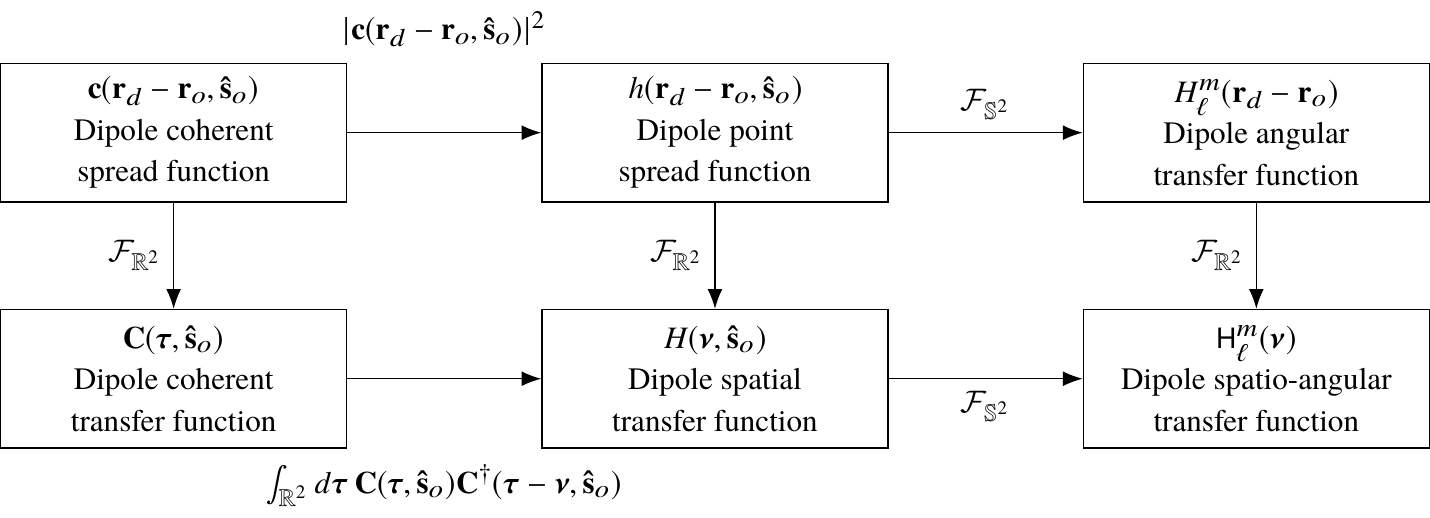}
  \caption{There is one transfer function for each set of object-space basis
    functions, and these transfer functions are related by two-dimensional and
    spherical Fourier transforms---see center and right columns. There is an
    additional pair of coherent transfer functions that are useful for
    calculating the transfer functions---see left column.}
   \label{fig:transfer-functions}
 \end{figure}
    
We can plug Eq. \eqref{eq:absquare2} into Eq. \eqref{eq:dstf} and use the
autocorrelation theorem to rewrite the dipole spatial transfer function as
\begin{align}
  H(\bv, \so) = \int_{\mbb{R}^2}d\taup\,\mb{C}(\taup, \so)\mb{C}^\dagger(\taup - \bv, \so), 
\end{align}
where we have introduced the \textit{dipole coherent transfer function}
$\mb{C}(\taup, \so)$ as the two-dimensional Fourier transform of the dipole
coherent spread function:
\begin{align}
  \mb{C}(\taup, \so) = \int_{\mbb{R}^2}d\mb{r}\, \mb{c}(\mb{r}, \so)\,\text{exp}[-2\pi i\mb{r}\cdot\taup].
\end{align}
Physically, the dipole coherent transfer function corresponds to the
vector-valued electric field created by a dipole oriented along $\so$ in a
Fourier plane of the detector with appropriate scaling. Similar to the monopole
case, we can calculate the dipole-orientation-dependent fields in a Fourier
plane of the detector, scale appropriately to find the dipole coherent transfer
function, then use the relationships in Fig. \ref{fig:transfer-functions} to
calculate the other transfer functions. Finally, we note that the dipole
coherent transfer function is identical (up to scaling factors) to what
Agrawal et. al. call the \textit{Green's tensor} \cite{agrawal2012} and Novotny
and Hecht's \textit{dyadic point spread function} multiplied by the dipole
moment vector \cite{nov2006}.

\section{Discussion}\label{sec:discussion}
\subsection{When are dipole transfer functions necessary?}
Model mismatch can lead to biased estimates of the position and orientation of
fluorophores, so the most accurate fluorescence microscopy experiments will
always use dipole transfer functions over monopole transfer functions. However,
in many practical situations noise will mask the effects of vector optics and
dipoles. If the fluorophores are rotationally unconstrained or there are many
randomly oriented fluorophores in a resolvable volume---common situations in
biological applications---then the effects of vector optics and dipoles will be
masked by noise in all but the highest-SNR regimes. Therefore, the dipole
transfer functions are most broadly useful when the sample contains fluorophores
that are rotationally constrained. As rotational constraints increase, the
effects of vector optics and dipoles will become apparent in lower SNR regimes.
In the next paper of this series we will calculate the dipole transfer functions
for a 4-$f$ imaging system and investigate the conditions under which the
monopole and dipole models are identical.
 
\subsection{Alternative transfer functions}
Throughout this work we have used the spherical harmonic functions as a basis
for functions on the sphere, but there are other basis functions that can be
advantageous in some cases. Several works \cite{aguet2009, backer2014,
  brasselet2011, zhang2018, zhang2018-2} have used the second moments as basis
functions for the sphere because they arise naturally when computing the dipole
point spread function. These works use an alternative to the dipole angular
transfer function that uses the second moments as basis functions so the forward
model can be written as
\begin{align}
  g(\rd) = \sum_{j=1}^6 \int_{\mbb{R}^2}d\ro\, H_j(\rd - \ro)F_j(\ro),
\end{align}
where
\begin{align}
  H_j(\rd - \ro) &= \int_{\mbb{S}^2}d\so\, h(\rd - \ro, \so)Z_j(\so),\\
  F_j(\ro) &= \int_{\mbb{S}^2}d\so\, f(\ro, \so)Z_j(\so),
\end{align}
and $Z_j(\mh{s}) = \{s_x^2, s_y^2, s_z^2, s_xs_y, s_ys_z, s_xs_z\}$ are the
second moments. This formulation is similar to the dipole angular transfer
function approach because it can exploit the spatial sparsity of the sample, but
it does not require a cumbersome expansion of the dipole point spread function
onto spherical harmonics.

However, the spherical harmonics provide several advantages over the second
moments. First, the spherical harmonics form a complete basis for functions on
the sphere, while the second moments span a much smaller function space. The
usual approach to extending the span of the second moments is to use the fourth
(or higher) moments, but this extension requires a completely new set of basis
functions while the spherical harmonics can be extended by simply adding higher
order terms. Second, the spherical harmonics are orthonormal, which will allow
us to deploy invaluable tools from linear algebra---linear subspaces, rank, SVD,
etc.--- to analyze and compare microscope designs. Finally, using the spherical
harmonics provides access to a set of fast algorithms. The naive expansion of an
arbitrary discretized $N$ point spherical function onto spherical harmonics (or
second moments) requires a $\mc{O}(N^2)$ matrix multiplication, while pioneering
work by Driscoll and Healy \cite{driscoll1994} showed that the forward discrete
fast spherical harmonic transform can be computed with a $\mc{O}(N(\log N)^2)$
algorithm and its inverse can be computed with a $\mc{O}(N^{3/2})$ algorithm. To
our knowledge no similarly fast algorithms exist for expansion onto the
higher-order moments.

Zhenghao et. al. \cite{zhanghao2017} have used the circular harmonics to model
the orientation of dipoles. The circular harmonics are complete and orthogonal,
but they artificially restrict the reconstructed dipoles to the transverse
plane of the microscope---a rare situation in real experiments.

The diffusion magnetic resonance imaging community uses both the second moments
(or second-order tensor) basis functions \cite{basser1994} and the spherical
harmonic basis functions \cite{tournier2004}. Descoteaux et. al. have provided
an explicit transformation matrix to convert between these basis functions
\cite{descoteaux2006}.

\subsection{Towards spatio-angular reconstructions}
We have focused on modeling the mapping between the object and the data in this
paper, but ultimately we are interested in reconstructing the object from the
data. Applying the monopole approximation simplifies the reconstruction problem
because both object and data space are $\mbb{L}_2(\mbb{R}^2)$, so we can
directly apply regularized inverse filters and maximum likelihood methods. The
dipole model expands object space to $\mbb{L}_2(\mbb{R}^2\times \mbb{S}^2)$, so
the inverse problem becomes much more challenging. In future work we will use
the singular value decomposition to find inverse filters, and we will consider
using polarizers and multiple views to increase the size of data space. 

\section{Conclusions}
Many models of fluorescence microscopes use the monopole and scalar
  approximations, but complete models need to consider dipole and vector optics
  effects. In this work we have introduced several transfer functions that
  simplify the mapping between the dipole density and the irradiance pattern on
  the detector. In future papers of this series we will calculate these transfer
  functions for specific instruments and use the results to simulate and analyze
  data collected by these instruments.

\section*{Funding}
National Institute of Health (NIH) (R01GM114274, R01EB017293).

\section*{Acknowledgments}
We thank Kyle Myers, Harrison Barrett, Scott Carney, Luke Pfister, Jerome Mertz,
Sjoerd Stallinga, Mikael Backlund, Matthew Lew, Min Guo, Yicong Wu, Shalin
Mehta, Abhishek Kumar, Peter Basser, Marc Levoy, Michael Broxton, Gordon
Wetzstein, Hayato Ikoma, Laura Waller, Ren Ng, Tomomi Tani, Michael Shribak, Mai
Tran, Amitabh Verma, Xiaochuan Pan, Emil Sidky, Chien-Min Kao, Phillip Vargas,
Dimple Modgil, Sean Rose, Corey Smith, Scott Trinkle, and Jianhua Gong for
valuable discussions during the development of this work. TC was supported by a
University of Chicago Biological Sciences Division Graduate Fellowship, and PL
was supported by a Marine Biological Laboratory Whitman Center Fellowship.
Support for this work was provided by the Intramural Research Programs of the
National Institute of Biomedical Imaging and Bioengineering.

\section*{Disclosures}
The authors declare that there are no conflicts of interest related to this article.

\appendix
\section{Spherical harmonics and the spherical Fourier transform}\label{sec:sph}
The spherical harmonic function of degree $\ell$ and order $-\ell \leq m \leq \ell$
is defined as \cite{schaeffer2013}
\begin{align}
Y_\ell^m(\vartheta, \varphi) = \sqrt{\frac{2\ell+1}{4\pi}}\sqrt{\frac{(\ell-|m|)!}{(\ell+|m|)!}}P_\ell^m\left(\cos\vartheta\right)\exp(i m \varphi),
\end{align}
where $P_\ell^m(\cos\vartheta)$ are the associated Legendre polynomials with the
Condon-Shortley phase
\begin{align}
  P_\ell^m(x) = (-1)^m(1-x^2)^{|m|/2}\frac{d^{|m|}}{dx^{|m|}}P_\ell(x),
\end{align}
and $P_\ell(x)$ are the Legendre polynomials defined by the recurrence
\begin{align}
  P_0(x) &= 1,\\
  P_1(x) &= x,\\
  \ell P_\ell(x) &= (2\ell-1)xP_{\ell-1}(x) - (\ell-1)P_{\ell-2}(x). 
\end{align}
The spherical harmonics are orthonormal, which means that
\begin{align}
  \int_{\mbb{S}^2}d\mh{s}\, Y_\ell^m(\mh{s}){Y}_{\ell'}^{m'*}(\mh{s}) = \delta_{\ell\ell'}\delta_{mm'},
\end{align}
where $\delta_{\ell\ell'}$ denotes the Kronecker delta. The spherical harmonics form a
complete basis, so an arbitrary function on the sphere $f(\mh{s})$ can be
expanded into a sum of weighted spherical harmonic functions
\begin{align}
  f(\mh{s}) = \sum_{\ell=0}^{\infty}\sum_{m=-\ell}^{l}F_\ell^mY_\ell^m(\mh{s}).
\end{align}
We can find the spherical harmonic coefficients $F_\ell^m$ for a given function
using Fourier's trick---multiply both sides by $Y_\ell^{m*}(\mh{s})$,
integrate over the sphere, and exploit orthogonality to find that
\begin{align}
  F_\ell^m = \int_{\mbb{S}^2}d\mh{s}\, f(\mh{s})Y_\ell^{m*}(\mh{s}).
\end{align}
The coefficients $F_\ell^m$ are called the \textit{spherical Fourier transform}
of a spherical function.

\end{document}